# Molecular Programming Pseudo-code Representation to Molecular Electronics

Manas Ranjan Pradhan, E.G.Rajan

**Abstract**— This research paper is proposing the idea of pseudo code representation to molecular programming used in designing molecular electronics devices. Already the schematic representation of logical gates like AND, OR, NOT etc.from molecular diodes or resonant tunneling diode are available. This paper is setting a generic pseudo code model so that various logic gates can be formulated. These molecular diodes have designed from organic molecules or Bio-molecules. Our focus is on to give a scenario of molecular computation through molecular programming. We have restricted our study to molecular rectifying diode and logic device as AND gate from organic molecules only.

———————————— ◆ ————————————

## 1 INTRODUCTION

Programming in general we mean the structuring of a particular activity in terms of algorithm or sequence of pseducode.This pseudo code used through computer programming or system programming language to verify the results. While designing molecular logic device from molecular diodes, the structure of the organic molecules also changes. This change in structure of the molecules affects the computation factors which are nothing but the flow of electrons. The target results with various voltage flows also get on change with flow of electrons. Designing molecules and molecular systems are like programming electronic devices. Molecular programming model gives a designing of molecular systems with information processing capability using molecular programming. It aims at establishing systematic design principles for molecules and molecular systems with information processing capabilities and developing methods to ease their design and construction [6]. Methods in molecular programming are roughly classified in to those for designing molecules and those designing molecular reactions. [5] We taken the computation paradigms with donor –acceptor-insulator as a single molecule platform and the interactions among molecules taking in to consider the arrangement of molecule according to the structure.

## 2 BASIC ARCHITECTURE FOR MOLECULAR PROGRAMMING

The architecture of the processes involved in giving a model to molecular designing of molecular device shown in fig1.The process starts with choosing the organic molecules to design molecular diode and extend to molecular logic gates(We have taken only AND). The proper simulation of molecules can be found out with choosing the donor, acceptor and insulator and their arrangement according to the re-sidue molecule attached on it. The molecular properties get studied taking Donor-Insulator-Acceptor as single molecule with benzene as nuclei. The molecular orbital properties also studied with many molecular orbital combinations with donor, acceptor and insulator as separate entity flowing along with benzene. The overlapping properties of highest orbitals found out once the multi-nuclei molecules get arranged with each other arbitrarily in the same surface dominated by benzene. The evaluation of the electronic properties has done with orbital motion and energy studies have done by taking computational chemistry and molecular physics rules.

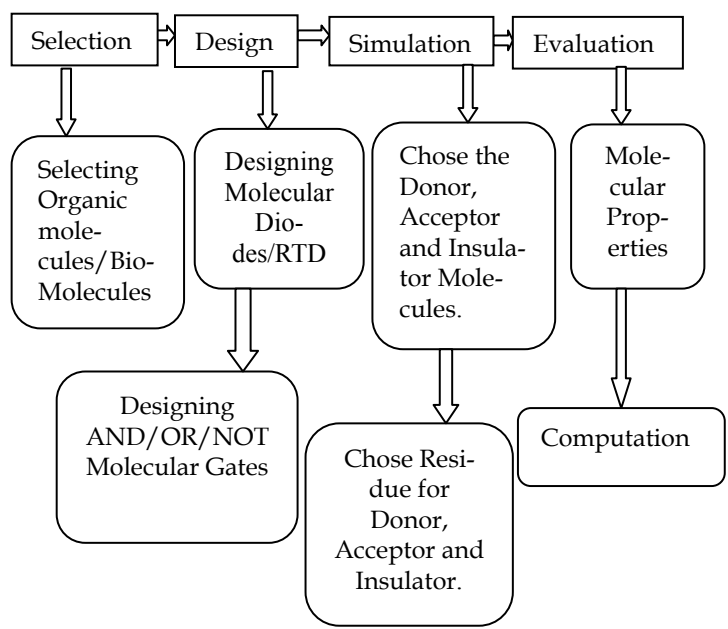

**Fig1**.Architecture of Molecular Simulation and Programming

• *Manas Ranjan Pradhan , is with IBS,Hyderabad,India.*
• *E.G.Rajan , is with MGNIRSA,Hyderabad,India.*

## 3 MOLECULES STUDY

Molecular systems, or systems based on organic molecules, possess interesting and useful electronic properties. Going through various family of organic molecules and determine the electrical and insulating properties of molecules is the primary need. In this regard we have studied organic molecules.

*Organic Molecules as Conductors and Insulators*

Two primary molecules that could be used as potential basis for current-carrying molecular-scale electronic devices are polyphenylene chain and carbon nanotube [1].We primarily concern about polyphenylene chain type of molecules. So we consider conjugate aromatic organic molecules as conductors or wires. We have considered the aliphatic organic molecules as insulators. The aromatic compound which acts as current carrying capabilities is based on benzene. The aliphatic group which acts insulators does not contain benzene rings. Aliphatic compounds can be cyclic, like cyclohexane, or acyclic, like hexane. They also can be saturated, like hexane, or unsaturated, like hexane. In aliphatic compounds, carbon atoms can be joined together in straight chains, branched chains, or non-aromatic rings (alicyclic). They can be joined by single bonds (alkanes), double bonds (alkenes), or triple bonds (alkynes). Besides hydrogen, other elements can be bound to the carbon chain, the most common being oxygen, nitrogen, sulfur, and chlorine. The simplest aliphatic compound is methane ($CH4$).Aliphatics include alkanes (e.g. paraffin hydrocarbons), alkenes (e.g. ethylene) and alkynes (e.g. acetylene).

## 4 DESIGNING MOLECULAR DIODES & AND GATE

The molecular diode designed either in form of a rectifying diode or in form of a resonant tunneling diode. A molecular diode contains two terminals and functions like a semiconductor p-n junction and has electronic states as conductive state (ON) and less conductive state (OFF).Aviram and Ratner have suggested that electron donating constituents make conjugated molecular groups having a large electron density (N-type) and electron withdrawing constituents make conjugated molecular groups poor in electron density (P-type). According to them, a non-centro symmetric molecule having appropriate donor and acceptor moieties linked with a  -bridge (may act as a insulator or internal register) and connected with suitable electrodes will conduct current only in one direction - acting as a rectifier. In this analysis of Aviram and Ratner in a *D- -A* molecule, the lowest unoccupied molecular orbital (LUMO) and highest occupied molecular orbital (HOMO) can be aligned in such a way that electronic conduction is possible only in one direction making it function like a molecular diode[1].

The structure of the mono-molecular rectifying diode proposed in [2] is shown in Fig. 2 and its schematic representation is shown in Fig. 3.

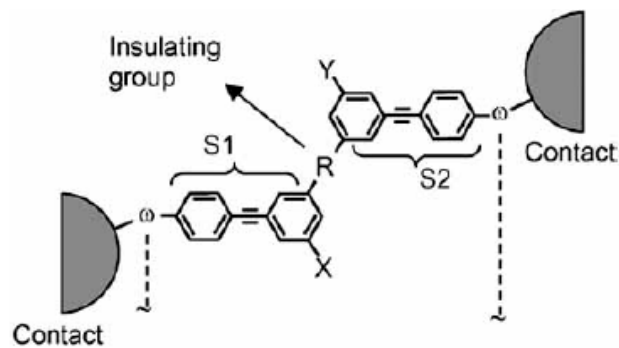

**Fig.2** Structure of Rectifying Molecular Diode (*monomolecular diode*).

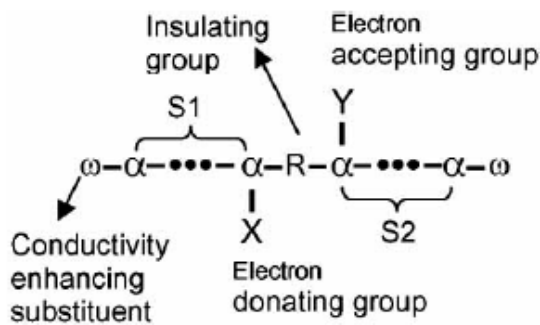

**Fig.3** Schematic Representation of rectifying Molecular Diode.

The circuit representation of a diode logic AND gate is shown in (Fig. 4) and the schematic representation of the Diode logic molecular AND gate is shown in (Fig. 5). The schematic of the exemplary poly-phenylene diode logic Molecular AND gate is shown in (Fig. 6).

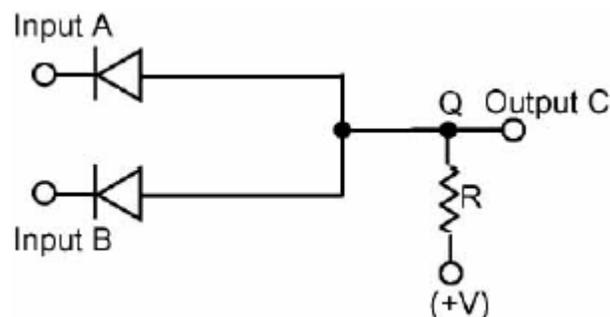

**Fig.4** Circuit Diagram of AND gate

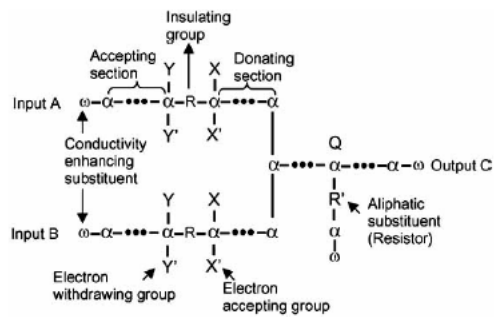

**Fig.5** Schematic Representation of AND gate

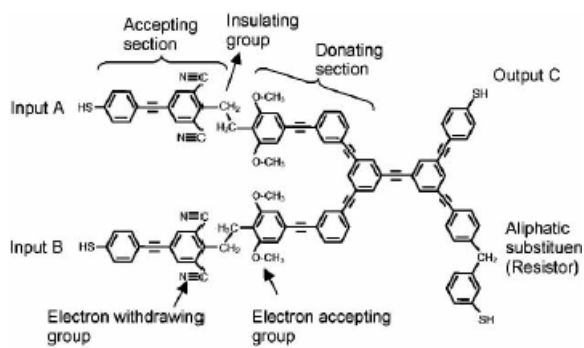

**Fig.6** Schematic of the exemplary poly-phenylene diode logic

## 5 SIMULATING

We have discussed in Part-3, the various molecules from organic molecules and biomolecules that can act as conductor and insulator. The choosing of Donor, Insulator and Acceptor depend on the electrical conductive of these molecules. Referencing Fig.2 we have used S1 as donate group and S2 as acceptor group in a single molecule.S1 and S2 are typically benzene constitute molecule. In a polyphenylene chain all these benzene molecules form chain and the electrical conductivity comes with donor group X and acceptor group Y. Referring to Fig.3 the contact metal(w) may taken any inorganic metal like Gold,Uranium,aluminium etc.The electrical properties and flow generally depict out if the valence electrons available for the molecule.

The Benzenes ring typically referred as C6H6 chemical formula. The combination of all these benzenes typically forms a polyphylene chain.

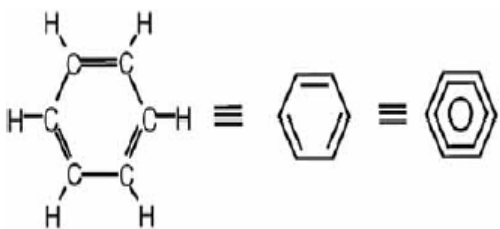

**Fig7.** A benzene structure

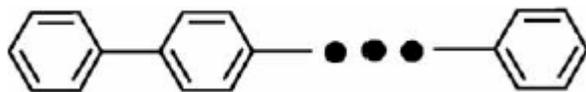

**Fig.8.** Polyphenylene chain

Benzene is a flat ring of SP² hybrid carbon atoms with their unhybridized p orbitals all aligned and overlapping. The conjugation and delocalization of the electrons in benzene gives this compound greater stability than non-conjugated cycles [1]. The double bond act as Pi-Bond and single bond as Sigma bond. The typical angle between hydrogen and carbon atom is 120 degree and bond length between carbon to carbon is 1.397 Armstrong. The Phenyl group formed by reducing one hydrogen atom and phenyl form by reducing 2 hydrogen atoms. The hydrogen atoms that getting on reduced in a polyphenylene chain typical adjusted with either a donating group ele-ment or molecule(X) or with acceptor group element(Y).Similar as in part 3 we have taken aliphatic molecules as insulator. The insulator typically an alkyl or acetylene functional group of organic family.

The donating group can be (X) = **(-NH2, -OH, -CH3, -CH2CH3, etc.)**
The different types of Acceptor residue that we take as (Y) = **(-NO2, -CH, -CHO, - NC etc.)**
The different types of Insulator that we take as (R) = **(-CH2-,-CH2-CH2-, etc.)**
The combination of X, Y and R in molecular conduction creates an uncertainty of electron flow.
The uncertainty arrives as

a) Fixing one X (say –NH2), we may have varieties of Y ((-NO2, -CH, -CHO, - NC) and (R) = (-CH2-,-CH2-CH2-)
b) Fixing one R (say –CH2), we may have varieties of Y ((-NO2, -CH, -CHO,- NC) and (X) = (-NH2, -OH, -CH3, -CH2CH3)
c) Fixing one Y (say –NO2), we may have varieties of (X) = (-NH2, -OH, -CH3, -CH2CH3) and (R) = (-CH2-,-CH2-CH2-)

Multiple donor/acceptor sites can be incorporated to adjust R.As shown in fig.9, the three potential barriers - one belongs to the insulating group and other two belongs to the contact between the molecule and the electrode on both sides in a zero biased condition [1]. The occupied energy levels in the metal contacts and the Fermi energy level $E_F$ are also shown. The pi-type energy levels (HOMO as well as LUMO) are elevated on the left of the central barrier due to the presence of the electron donating group X and similarly on the right of the central barrier the energy levels are lowered due to the presence of the electron withdrawing group Y. This causes energy difference $E_{LUMO}$ across the barrier. The electrons must overcome the potential barrier from electron acceptor doped section (S2) to electron donor doped section (S1) for current to flow and this forms the basis for the formation of the mono-molecular rectifying diode. The energy band diagram under forward bias conditions (left hand contact at higher potential than the right hand contact) is shown in Fig. **9**. Here, electrons are induced to flow by tunneling through the three potential barriers from right to left causing a forward current flow from left to right.

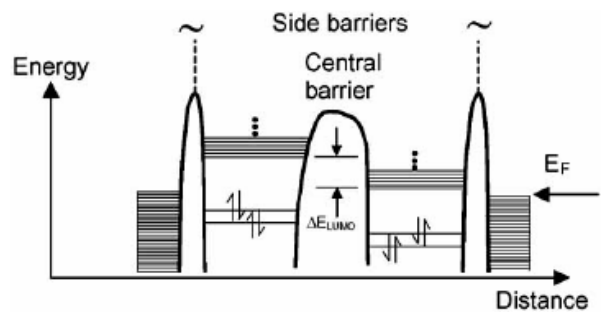

**Fig. (9).** Orbital energy diagram of the polyphenylene

monomolecular
rectifying diode under zero bias conditions [2].

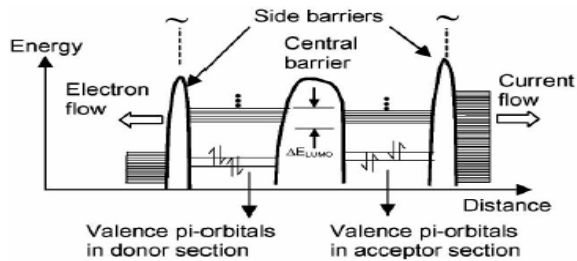

**Fig. (10)**. Orbital energy diagram of the polyphenylene monomolecular rectifying diode under forward bias conditions [2].

The situation is getting more complex if we analyze the situation Fig.5 and Fig.6 where two input of molecule string used but acting as single molecule of operation. Both of the molecular diodes are joined together forming a common node Q to which an aliphatic chain can be attached. Another conducting molecular wire is for the output C. This completes the formation of the larger single molecule [1]. Standard insertion and substitution techniques or other techniques can use for this molecular diode AND logic in a single molecule.

# 6 EVALAUTING

## 6.1 Molecualr Properties

As we know the various families of organic molecules are alkanes, alkenes, alkynes, alkadienes, cyclic carbohydrates,chiral compounds, aromatic compounds, alchohols & ethers, aldehyde & ketones, carboxylic, acids & esters, amines & amides and polymores. Out of all these all are not useful for molecular electronics. Some are used for molecular electronics by combining with other elements of organic molecules and inorganic metals. We have extracted the molecular compounds which are responsible for electrical properties act as conductors and other molecules which act as insulators. Benzene is a flat ring of sp2 hybrid carbon atoms with their unhybridized p orbitals all aligned and overlapping. To design a plophenylene based molecules we consider benzene structure as $C_6H_6$, phenyl group as $C_6H_5$, phenylene group as $C_6H_4$. Combining so many plophenylene we design plophenylene chains.
Aliphatics include alkanes, alkenes (e.g. ethylene) and alkynes (e.g. acetylene).The smallest alkyl group is methyl, with the formula $CH_3$. Alkyls form homologous series with general formula $C_nH_{2n+1}$. Alkyls include methyl, $CH_3$· (named after methane), ethyl ($C2H5$·), propyl ($C3H7$·), butyl ($C4H9$·), pentyl ($C5H11$·), and so on. Alkyl groups that contain one ring have the formula $C_nH_{2n-1}$, e.g. cyclopropyl and cyclohexyl. We design (-CH2-) Methylene, (-CH2-CH2-) Di-methylene and a chain like (-CH2 -CH2-CH2--CH2--) forms alkyl chains act as insula-tors. Some general examples of aliphatic molecules are ethane, isobutene and acetylene.
The system itself now seems to more complex in studying the molecular properties and atomic properties for electron flow as types of X, Y and R having different chemical, physical, and electronic properties. These compounds are made up of molecules whose atoms bind to one another through "covalent" bonds. The current carrying nature with energy variance now gives it a complex model. Studying the energy levels also complex as the electron flows in a particular potential barrier is quite uncertain. The question arises whether Heisenberg principle or Schrodinger equation will give a solution to this many electrons, many atoms and many molecules acting as a single molecule, responsible for current flows. The atomic properties studied by studying the quantum mechanical model of the atom with electrons occupy orbitals and volumes of space around the nucleus with a high probability of finding the electron.ln this case benzene and X-in donating section will create a molecular nucleus of their own with a probability that the valence electrons of each atoms is now overlapping with each other. The discrete energy level in each of these molecular interaction or arrangement is getting on disturbed. There is a significant probability of finding an electron of an atom at any position within a spherical volume surrounding the nucleus of each atom of benzene as well each atom of X.The factors that influence the atomic properties are the shape of the molecule (linear, trigonal planar, tetrahedral, trigonal bipyramidal, octahedral etc.),the bond angle,a lone pair of electrons is a non-bonding pair of electrons, atomic orbitals and their shapes(s, p, d, f), spin quantum number, many electron atoms (electronic configuration of atoms),molecular structure (VSEPR model),covalent bond (Valence Bond theory-Molecular Orbital Theory, Orbital overlap, Directionality of bonds & hybridization(s & p orbitals only ), resonance ( orbital theory, Orbital energy level diagram, bond order) and Intermolecular Forces( polarity; dipole moments).

The quantum mechanical property of molecule as whole makes us to know how to find the energy states of the molecule or subsidiaries that used for molecular electronics. Chemical bonds are a source of energy and the movement of molecules in space is kinetic energy. The vibrations and rotations of molecules is another source of chemical energy along with the chemical reaction which is a rearrangement of atoms.
Erwin Schrodinger developed the equation which is used today to understand atoms and molecules. Erwin Schrodinger derived ideas from de Broglie, Heisenberg and others and put them together in a single equation. Solving this equation can in principle predict the properties and reactivities of all atoms and molecules. The general Schrodinger equation represent as $\mathbf{H\Psi = E\Psi}$, where H=Hamiltonian $\Psi$=Orbital and E=Energy. Orbitals and Energies are the central objects that determine the properties of atoms and molecules in the new Quantum Theory.

Although the Schrodinger equation is too difficult to solve for any but the simplest atoms/molecules, we can nevertheless extract some essential conclusion from it-like Energies are quantized, the orbitals, associated with each energy, determine where the electrons are located. The Aufbau Principle rules some time applied which discuss about quantum numbers, shells, subshells and orbitals.

## 6.2 Molecualr Programmming pseudo code

*Start:* ******
Approach: Top-Down: Macromolcule level to atomic level study

**Process1:** *Selection*

Step 1.Chose organic molecules
Step2: Define the molecule from functional group (Aromatic Compounds-Benzene and its derivatives)
Step3: Associate a chain of Benzene molecules

**Process2:** *Designing:*

*Part-1: Molecular Diode*
 Step 1. Associate Benzene molecules with other functional groups
 Step 2. Chose functional group from Hydroxyl, Carbonyl, Carboxyl, Amino, Sulfhydryl, Phosphate
 Step 3. Structure the Diode in Donor-Insulator-Acceptor part
 Step 4. Design the donor part by associating functional group element with benzene chain
 Step 5. The functional group taken in the donor part in association with benzenes denoted as
  X: (-NH2, -OH, -CH3, -CH2CH3…etc.) (Extracting from step 2: Designing)
 Step 6. The functional group taken in the Acceptor part in association with benzenes denoted as
  Y: (-NO2, -CH, -CHO, - NC) ….etc. (Extracting from step 2: Designing)
 Step 7. Chose element from Alkyl group: Aliphatic Insulators Bridge (Non-Aromatic Compound)
   Internal Register  R:  (-CH2-,-CH2-CH2- etc…)
 Step 8. Join Carbon atoms in insulator by single bonds (alkanes), double bonds (alkenes), or triple bonds (alkynes).
 Step 9. Join carbon atoms in insulator by:  hydrogen, oxygen, nitrogen, sulfur, and chlorine.
 Step 10.Put contacting element in donor and acceptor site by conducting inorganic metal as gold, uranium, aluminium
 Step 11. Substitue element w can put from functional group step 2

*Part2:  AND gate*
 Step 1.Take two inputs say A & B: Two molecular diode describe in part 1
 Step 2.Out put of (A & B) let say C
 Step 3.Describe the X, R, Y of A & B
 Step 4.Design the common node for A & B for electron conductivity
 Step 5.Put another external Resistance Load (R')
 Step 6.Output detect in Point C
 Step7.Chosen from any functional group linking with benzene
 Step8.Choosing the donating section along with X and benzene chain in Input A
 Step 9.Choosing the accepting section along with Y and benzene chain in Input A
 Step 10.Choosing the accepting group along with X and benzene chain in Input B
 Step 11.Choosing the withdrawing group along with Y and benzene chain in Input B
 Step 12.Substitute element w can put from functional group step 2-*Part 1*

**Process3:** *Simulating*

*Part 1.Finding donor properties of X*
 Step1:  Electronic configuration of X
 Step2: Label molecular orbitals according to symmetry
 Step3: Find out σ and π orbitals
 Step4: Electron affinity of X
 Step 5: Calculate molecular mass
 *Part 2.Finding properties of Benzene Chain*
 Step1: Electronic configuration of benzene chains
 Step2: Label molecular orbitals according to symmetry
 Step3: Find out σ and π orbitals
 Step 4: Electron affinity of benzene chains
 Step 5: Calculate molecular mass
*Part 3.Finding Accepting properties of Y*
  Step1:  Electronic configuration of Y
 Step2:  Label molecular orbitals according to symmetry
 Step 3: Find out σ and π orbitals
 Step4: Electron affinity of Y
 Step5: Calculate molecular mass
 *Part 4.Finding Insulating properties of R*
   Step 1: Electronic configuration of Insulating chain
   Step 2: Label molecular orbitals according to symmetry
   Step 3: Find out σ and π orbitals
   Step 4: Electron affinity of R
   Step5: Calculate molecular mass
 Part 5: *Structure for Donor-Insulator-Acceptor configuration*
    Step1: Chose benzene as the central nucleus of the Donor-Insulator-Acceptor set (single molecule)
    Step2: Chose the Set D-I-A depending on X, Y, R have taken along with w

Step3: Checking the overlapping of molecular orbitals

Step 4: Repeat the process 3 for different X, Y, R, and w

Step 5: Study the probability of orbital motion

**Process4:** *Evaluation*

*Part 1-Molecular Properties study*

Step1: Checking the overlapping in molecular orbitals of X, Y and R

Step2: Each of molecular orbital form atomic orbital

Step 3: Each electrons of X, Y, R molecular orbits move under the influence of any molecule of X, Y and R.

Step4: **Set Constraints every electrons of X, R, Y molecules move under the influence only benzene Nuclei.**

Step5: Assumed that the molecular orbital wave function ($\psi_j$) may be written as a simple weighted sum of the n constituted atomic orbitals Xi, Yi and Ri.

$$\psi_J = \sum_{i=1}^{n} C_{ij} (X_i, Y_i, R_i)$$

The Cij coefficients may be determined numerically by substitution of this equation into the Schrödinger equation and application of the variational principle. This method is called the linear combination of atomic orbitals approximation.

Step6: Finding the probability of orbital motion:

It is called bonding orbital if orbital have a higher probability of being between nuclei than elsewhere and hold the nuclei together. It is anti-bonding orbital if the electrons spend more time elsewhere than between the nuclei of X, Y, R and benzene and weaken the bond. Electrons in non-bonding orbitals tend to be almost atomic orbitals (Some cases w) neither contribute nor detract from bond strength.

In Molecular Orbital theory, any electron in a molecule may be found anywhere in the molecule, since quantum conditions allow electrons to travel under the influence of an arbitrarily large number of nuclei, so long as permitted by certain quantum rules.

Step 7: Check the rules:

i) Pauli Exclusion Principle, two electrons cannot share the same set of quantum numbers within the same system.

ii) Hund's rule for Ground state energy calculation

iii) The Aufbau principle

*Part 2: Computation*

Step1: Finding energy state:

Each molecular orbit has fixed amount of energy with it and get changes if the molecular association get on change. The outer orbit electrons in a molecule posses more energy than the innermost orbit. The configuration that corresponds to the lowest electronic energy is called the ground state. Any other configuration is an excited state.

Step 2: Calculating LUMO & HOMO

To identify the HOMO(Highest Occupied Molecular Orbital) and theLUMO(Lowest Unoccupied Molecular Orbital )of a given molecule one can find out all the molecular orbitals and fill them with the available electrons, or use a generic ordering of orbitals, and use valence bonding for σ-type and lone pair orbitals, and molecular orbitals for π-systems as an approximation[25].

The generic ordering of molecular orbitals (from highest energy to lowest energy): σ* - almost never occupied in the ground state, π* - very rarely occupied in the ground state, n - nonbonding (ie. lone pairs),π - always occupied in compounds with multiple bonds (Pi Bond),σ - at least one occupied in all molecules (Sigma Bond) and a - an empty orbital .The kinds of electrons present are identified through observation of the Lewis structure. Although this generic system will identify the correct HOMO and LUMO most of the time, there are exceptions to this classification. To identify any resonances in the system, the entire π system must be considered.

Repeat the step for other gates.

**End*******

# 7   DISCUSSIONS

*Discussion 1:Part1-process4- step 5: Molecular Orbital features of benzene*:

It is a hexagonal ring of 6 carbon atoms in which 24 of the 30 total valence bonding electrons are located in 12 σ (sigma) bonding orbitals which are mostly located between pairs of atoms (C-C or C-H).The remaining 6 bonding electrons are located in 3 π (pi) molecular bonding orbitals that are delocalized around the ring. Two are in an MO which has equal contributions from all 6 atoms. The other two orbitals have vertical nodes at right angles to each other. It is evident that the 3 molecular pi orbitals form a combination which evenly spreads the extra 6 electrons over 6 carbon atoms unlike valence bond [26]

Discussion 2: *Part1-process4-step5: Molecular Orbital features of methane* (Alkane series-X molecule in this model):

The 8 valence electrons are found in 4 Molecular Orbitals which are spread over all 5 atoms. It approximated 4 localized orbitals, similar in shape to sp3 hybrid orbitals predicted by Valence Bond theory. This creates σ (sigma) bonds, but not the π (pi) orbitals. However, the ionization and spectroscopic predictions can happen to the delocalized MO picture. Upon ionization of methane, a single electron is taken from the MO which surrounds the whole molecule, weakening all other 4 bonds equally in contrast to VB theory where one electron is removed for an sp3 orbital, resulting resonance between four valence bond structures, each of which has a one-electron bond.[26]

Discussion 3: *Step 2-Part 2-Process 4-Super molecular interaction energy:*

A straightforward approach for evaluating the interaction energy is to calculate the difference between the energies of isolated objects and their assembly. By calculating the energies for monomers, dimers, trimers, etc., in an N-object system, a complete set of two-, three-, and up to N-body interaction energies can be derived. The super molecular approach has an important disadvantage in that the final interaction energy is usually much smaller than the total energies from which it is calculated, and therefore contains a much larger relative uncertainty. A molecule or atom that has a positive electron affinity is often called an electron acceptor and may undergo charge-transfer reactions. The electron donating power of a donor molecule is measured by its ionization potential which is the energy required to remove an electron from the highest occupied molecular orbital.

The overall energy balance ($\Delta E$), i.e., energy gained or lost, in an electron donor-acceptor transfer is determined by the difference between the acceptor's electron affinity (A) and the ionization potential (I) as $\Delta E=A-I$.

Electronegativity is a chemical property that describes the ability of an atom or a functional group to attract electrons or electron density towards itself. The term "ionization energy" is sometimes used as a name for the work needed to remove (to infinity) the topmost electron from an atom or molecule adsorbed onto a surface. So, in the case of surface-adsorbed atoms and molecules, it is called as "electron binding energy". Both these names are also sometimes used to describe the work needed to remove an electron from a "lower" orbital (i.e., not the topmost orbital) to infinity, both for free and for adsorbed atoms and molecules and needed to specify the orbital from which the electron has been removed. The most general form is the time-dependent Schrödinger equation, which gives a description of a system evolving with time. For systems in a stationary state, the time-independent Schrödinger equation is sufficient. Approximate solutions to the time-independent Schrödinger equation are commonly used to calculate the energy levels and other properties of atoms and molecules [28].

## 7 Current and future development

As we have gone through the molecular programming approach to AND gate designed from Molecular Diodes, similar way we can go for other gates OR, XOR, NOT etc.The same principle can apply to if we are designing molecular gates from RTD.The difference will be in potential barrier rules that encounter much in case of RTD.Further to study we can use molecular programming to design molecular gates from DNA molecules which is more complex. Already people have tried with different biological process in deforming or constructing or designing the DNA sequence, but as for the molecular orbital approach are in DNA computing yet not accomplished. We have still to do more research on computational chemistry aspects to molecular orbitals that affects the electrical properties and we are in way to study more on Schrödinger equation for energy level study in molecular orbitals. The molecular programming approach from DNA molecules have already tried by different persons as DNA has more computing power and computational biology already gives a base to study on it. Lots of things need to consider for matching biological process to molecular programming for molecular electronics. The ultimate goal of molecular computing to molecular electronics is to synthesize artificial information processing systems at molecular level with new paradigm, algorithm and programming [4].The further studies to molecular orbitals for Polyatomic molecules, the use of Schrödinger equation and Density Functional Theory(DFT) and Kopman's theorem application to energy band of molecular computation need to work out. The Born–Oppenheimer approximation application to molecular computation is to study to find amore robust picture of molecular orbitals that will useful for molecular electronics.

## Acknowledgment

The authors wish to thank Dr.Swaminathan Research Foundation, MGNIRSA.Hyderabad, India and pentgram research centre, Hyderabad, India and IBS, Hyderabad for literature survey and research findings.

**Manas ranjan pradhan** Mr.Manas Ranjan Pradhan received M.Sc (Physics-X-rays) from National Institute of Technology, Rourkela, India and M.Tech (Computer Science) from Utkal University; Bhubaneswar, India .He is currently doing his PhD on Molecular Computing under University of Mysore, India. Presently he is working as assistant professor in IBS, Hyderabad, India.His research interests are Bioinformatics, Speech recognition, and Nano computing.

**E.G.rajan** is an Electronics Engineer and a Professor of Signal Processing having about 30 years of experience in teaching, research and administration. He received his PhD degree in Electrical Engineering, (Signal and Image Processing), from Indian Institute of Technology (IIT), Kanpur, U.P., M.E. degree in Applied Electronics, from Madras University. His contribution to the state-of-the-art of Electronic Warfare has been well recognized in the Government and industrial sectors. He received Distinguished Scientist and Man of the Millennium Award from who is who Bibliographical Records, Cambridge, 2000. He isthe father of a novel paradigm Symbolic Computing .His Mathematical Transform, which goes by his name as Rajan Transform used for Pattern Recognition and Analysis purposes.Currentlyhe is the director of RGISIT,MGNIRSA,Hyderabad,India.